\documentclass[preprint,11pt]{aastex}

\usepackage{graphicx}
\usepackage{amsmath,amssymb}
\usepackage{natbib}
\usepackage{array}

\begin{document}

  \title{Observability of pulsar beam bending by the Sgr~A* black hole}
  \author{Kevin Stovall, Teviet Creighton, Richard H. Price, Fredrick A. Jenet
}
  \affil{Center for Gravitational Wave Astronomy and Department of Physics and
    Astronomy, University of Texas at Brownsville, Brownsville, Texas 78520}
  
  \begin{abstract}
    According to some models, there may be a significant population of
    radio pulsars in the Galactic center. In principle, a beam from
    one of these pulsars could pass close to the supermassive black
    hole (SMBH) at the center, be deflected, and be detected by Earth
    telescopes.  Such a configuration would be an unprecedented probe
    of the properties of spacetime in the moderate- to strong-field
    regime of the SMBH.  We present here background on the problem,
    and approximations for the probability of detection of such
    beams. We conclude that detection is marginally probable with
    current telescopes, but that telescopes that will be operating in
    the near future, with an appropriate multiyear observational
    program, will have a good chance of detecting a beam deflected by
    the SMBH.
  \end{abstract}
  \maketitle
  \section{Introduction}\label{sec:intro}
  
  Near-infrared observations of stars near our Galaxy's central
  supermassive black hole (SMBH) have revealed a larger number of
  young, massive stars than can be explained by typical star formation
  models~\citep{2005ApJ...620..744G, 2005ApJ...628..246E}. This
  ``paradox of youth"\citep{2005ApJ...620..744G} has pointed to
  the development of a possible continuing top-heavy initial mass
  function (IMF) in the region near the central 
  SMBH~\citep{2007ApJ...669.1024M, 2005MNRAS.364L..23N}. A top-heavy
  IMF near Sgr~A* would imply the existence of a large number of
  neutron stars in close proximity to the central SMBH. Current
  estimates suggest that there could be $\gtrsim 10^4$ neutron stars
  within 1\,pc of Sgr~A*~\citep{2005ApJ...622L.113M} and possibly
  $\sim1000$ pulsars within $\sim0.01$\,pc of Sgr
  A*~\citep{2004ApJ...615..253P}.

  Theoretical predictions have also suggested a cusp of
  $\sim20,000$ neutron stars in the central parsec of our
  Galaxy~\citep{2006JPhCS..54..252F,2006JPhCS..54..321H}.  X-ray
  observations have been consistent with this number of neutron
  stars~\citep{2007MNRAS.377..897D} but appear to have ruled out the
  presence of a larger number of neutron stars (e.g., 40,000). For
  this paper, we will take the results from
  \cite{2004ApJ...615..253P} and will assume the existence of 1000 pulsars
  within 0.01\,pc of Sgr~A*.  We will also follow Pfahl \& Loeb in
  assuming that $n(r)$, the density of pulsars as a function of
  distance from the Galactic center, falls off as $r^{-3/2}$, so that
  \begin{equation}\label{nvsr0}
    n=\frac{3}{8\pi}\times10^6\, {\rm pc}^{-3}\, (r/1\,{\rm pc})^{-3/2}\,.
  \end{equation}
 With
  this assumption it will turn out that the quantity of greatest importance to our
  predictions is the number of pulsars at distances from the central
  SMBH between 0.001\,pc and 0.01\,pc.

  In this paper we will consider the possibility that an appropriate
  program to monitor pulsar beams from Sgr~A* would detect a beam that
  is strongly deflected by the central SMBH.  Such a system would be
  of great interest, as precision timing of the radio pulses from such
  a system would measure the properties of the spacetime through which
  they propagate.  Preliminary work in the case of Schwarzschild
  black holes, presented in \cite{2009ApJ...697..237W} and
  \cite{2009ApJ...705.1252W} (hereafter Paper~I and Paper~II), has
  revealed a rich structure and multiplicity of pulses observed in
  such geometries; subsequent work will look at how pulse timing can
  be used to measure properties such as the mass and spin of the SMBH,
  and measure or constrain deviations in the higher multipoles of the
  spacetime from the predictions of general relativity.  In the
  present paper we will focus on determining the likelihood of
  observing a pulsar in such a configuration.  To do this we will first
  calculate the probability that a single pulsar, in orbit around the
  central SMBH, emits a signal in such a way that it is strongly
  deflected by the central SMBH, and reaches the Earth. From this we
  then infer the probability that the signal from one of the assumed
  number of pulsars is strongly deflected, reaches the Earth, and is
  detectable by  radio telescopes.
  
  The paper continues in Sec.~\ref{sec:bg} with a discussion of the
  model for the pulsar-SMBH system, and with assumptions about pulsar
  characteristics and telescope sensitivities. The heart of the paper
  is the method of computation of probabilities in
  Sec.~\ref{sec:prob}. Numerical estimates of probability, based on
  this approach, are given in Sec.~\ref{sec:results}, and considerations
  for an observing program are given in Sec.~\ref{sec:obsstrat}.  In
  Sec.~\ref{sec:conc} we conclude that an observing program, even with
  current radio telescopes, would have \emph{some} chance of detecting
  strongly-bent pulsar emissions, while later generations of
  telescopes will significantly increase the likelihood of observing
  these fascinating systems.

  \section{Background, model, and assumptions}\label{sec:bg}

  As a simplification in our probability estimates we take the SMBH in
  Sgr~A* to be a Schwarzschild hole. It is essentially certain, of
  course, that the SMBH is rotating, but the angular momentum $J$ is
  currently thought to be only about half of its maximum possible value of 
  $GM^2/c$ where $M$ is the mass of the SMBH
  \citep{MeliaBromleyEtal,2003Natur.425..934G}. Typically, the astrophysical properties of
  Kerr holes differ significantly from those of Schwarzshcild holes
  only when $J$ is very close to $GM^2/c$. Our preliminary
  investigations of Kerr holes, in work now underway, confirms
  this.
  While frame dragging by the black hole may slightly increase the
  probability of detecting beams bent in the prograde sense, and
  decrease the probability of retrograde-bent beams, these effects
  largely cancel when considering the population as a whole.  Thus the
  Schwarzschild approximation would appear to be justified for our
  purposes, in which it is the average properties that are of
  importance.

  \begin{figure}[h]
  \begin{center}
  \includegraphics[width=.4\textwidth ]{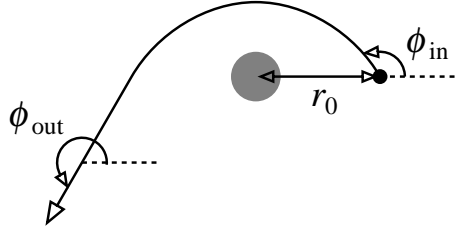}
  \caption{Photon trajectory and the definitions of the angles
    $\phi_{\rm in}$ and $\phi_{\rm out}=F(\phi_{\rm in};r_0)$.}
  \label{fig:photonorbit}
  \end{center}
  \end{figure}

  We will rely heavily on results in Papers~I and~II for pulsar beam
  deflection around a Schwarzschild hole. In those papers $\phi_{\rm
    in}$ is the angle between the direction of pulsar emission and the
  direction radially outward from the central SMBH at the emission
  event; the angle $\phi_{\rm out}$ is the angle between that same
  radial direction and the direction in which the pulsar beam is moving
  when it is asymptotically far from the SMBH,
  as sketched in Figure~\ref{fig:photonorbit}. In the absence of the
  bending of the beam, the two angles $\phi_{\rm in}$ and $\phi_{\rm
    out}$ would be equal. The effect of curvature of the beam is
  encoded in the function $F$ defined by
  \begin{equation}
    \phi_{\rm out}=  F(\phi_{\rm in};r_0)\,,
    \label{eq:F}
  \end{equation}
where $r_0$ is the distance of the emission point from the SMBH.
The computational method  for finding the $F$ function is 
discussed in Papers~I and~II. A practical approximation for $F$, 
useful for the considerations of this paper, is presented in the 
Appendix.

  \begin{figure}[h]
  \begin{center}
  \includegraphics[width=.7\textwidth ]{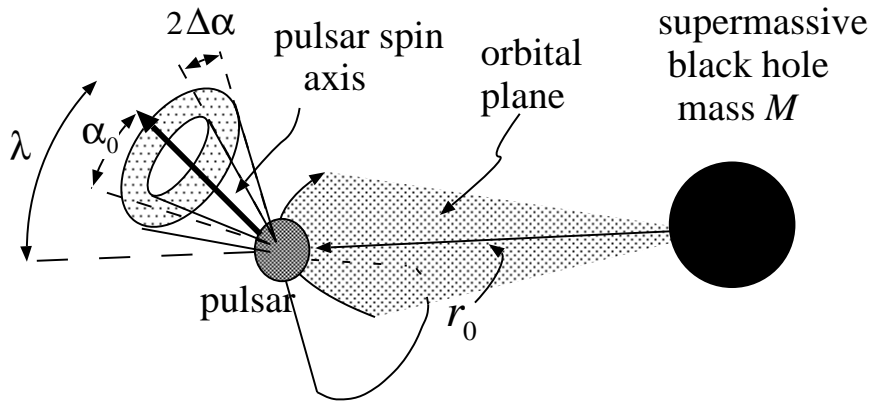}
  \caption{The geometry of the orbit, pulsar spin, and pulsar beam.}
  \label{fig:model}
  \end{center}
  \end{figure}

With the simplification to a nonrotating SMBH, we do not need to
consider any angle between the orbit of the pulsar and the spin axis
of the SMBH, The geometric parameters of interest are pictured in
Fig.~\ref{fig:model}.  The inclination of the \emph{pulsar} spin axis with
respect to the orbital plane is denoted $\lambda$; the beam of pulsar
emission is taken to have its center at angle $\alpha_0$ from the spin
axis, and to have width angular width $2\Delta\alpha$, so that the
pulsar emission is confined between conical surfaces with opening
angles $\alpha_0-\Delta\alpha$ and $\alpha_0+\Delta\alpha$, as shown
in Fig.~\ref{fig:model}.

In Fig.~\ref{fig:model}, $r_0$ denotes the radial distance of the pulsar from the 
SMBH at the moment of emission of a beam. We do {\rm not} assume circular oribits
in our probability calculations except in the calculations of orbital times in
Sec.~\ref{sec:results}.

We will assume that there is no favored alignment of the pulsar 
spin axis with the pulsar orbital plane, and will take $\lambda$
to be uniformly distributed over the sky.
Within a fraction of a pc from the central SMBH, we expect there to be
no significant alignment of the neutron star population with the disk
of the Galaxy, so we will take the orientation of the orbital plane also to
be randomly distributed over the sphere.

We will assume the angle $\alpha_0$, between the pulsar beam and the spin axis 
to be randomly distributed over the half sphere, and we assume that for every 
choice of $\alpha_0$ there is also a beam at $\pi-\alpha_0$. (In our probability 
estimates, we will avoid double counting in the case that $\alpha_0+\Delta\alpha>\pi/2$
and the conical regions of the two beams overlap.)

  \begin{figure}[ht]
  \begin{center}
  \includegraphics[width=.6\textwidth]{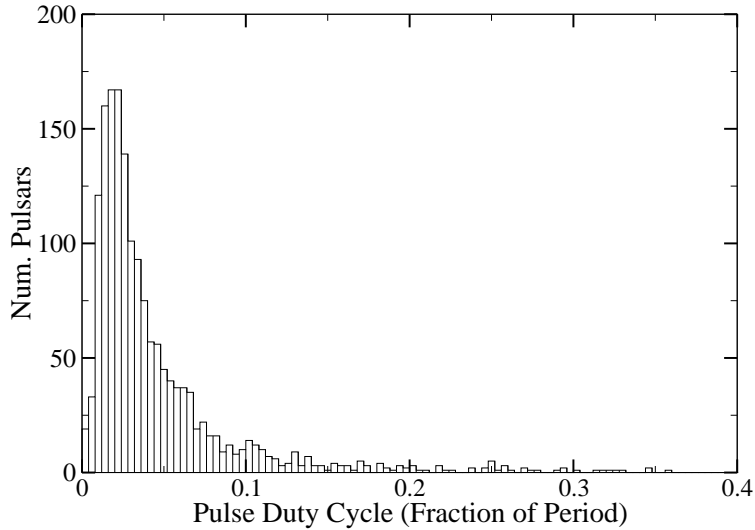}
  \end{center}
  \caption{ A histogram of the FWHM of currently known pulsars
  in units of fraction of duty cycle.}\label{fig:w50_hist}
  \end{figure}

The angular width $\Delta\alpha$ of the pulsar emission is a crucial
parameter in the probability of observation of deflected beams.
Figure~\ref{fig:w50_hist} shows a histogram, for currently
known pulsars,
of the FWHM of the pulse as a fraction of the pulse duty cycle. The mean of
this distribution is 4.6\%, so for probability calculations in this
paper we will use a duty cycle of 5\%, and therefore a \mbox{value of
9$^\circ$ for $\Delta\alpha$.}

\goodbreak

\begin{figure}[ht]
  \begin{center}
  \includegraphics[width=.6\textwidth]{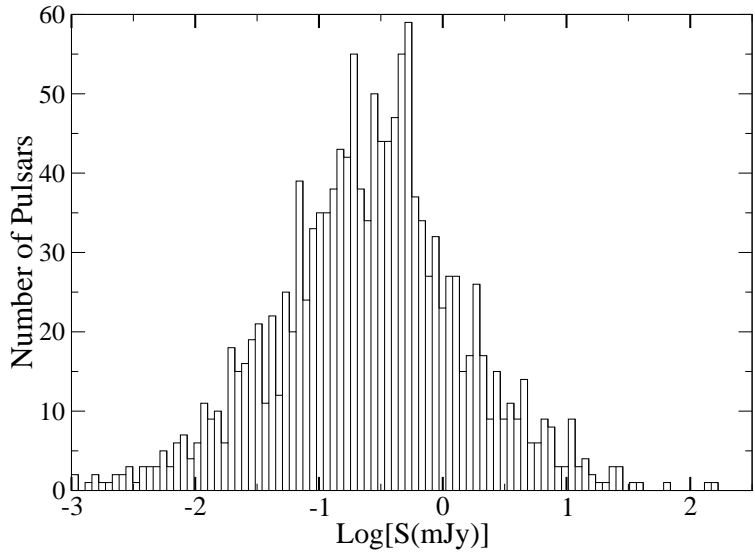}
  \end{center}
  \caption{
The distribution of pulsar L-band flux density $S$ that would be seen on Earth if all
cataloged pulsars were located at the center of the Galaxy.}
\label{fig:fluxgc} 
  \end{figure}

Of particular importance is the fact that a strongly bent beam will
generally be reduced in intensity in comparison with a directly
observed beam; thus our detection probabilities must account for the reduced
brightness of the source.
In terms of the bending function $F$ of Paper~I, the
``amplification'' factor (generally less than unity) for intensity is
given by
  \begin{equation}
  {\frac{I}{I_0}}=\frac{\sin\phi_{\rm in}}{\sin\left(F\right)\,({dF}/{d\phi_{\rm in}})}\,.
  \label{eq:ampfactor}
  \end{equation}
As a step in understanding how much reduction can be allowed if a pulsar 
beam is to be observed,  we use the radio luminosity at 1.4
  GHz\footnote{The available data are mostly for this band; however,
    observations of beams from Sgr~A$^*$ will likely need to be conducted in
    S band.  Nonetheless, we expect the distribution of luminosities to be
    roughly the same in the two bands.} for all pulsars for
  which this quantity has been calculated in
 the Australia Telescope
  National Facility \cite{ATNF}. 
We assume that the population of pulsars at the Galactic center has
the same distribution of luminosities  as these known pulsars. In that 
case, the distribution of pulsar flux densities $S$, observed at the Earth, 
would be that shown in   Fig.~\ref{fig:fluxgc}.

The minimum flux detectable at a telescope can be estimated with the following
equation from ~\cite{2004hpa..book.....L} (hereafter LK):
  \begin{equation}
  S_{\rm min}=\frac{({\rm SNR}_{\rm min})\beta_0 S_{\rm sys}}{\sqrt{n_{p}t_{obs}\Delta f}}\sqrt{\frac{W}{P-W}\;}\,.
  \label{eq:minflux}
  \end{equation}
Here $W$ and $P$ are respectively the pulsar pulse width and period.
From our assumption that $W/P$ is 5\% we get
  $\sqrt{W/(P-W)\;}=0.23$. The parameter $\beta_0$ is a correction
  factor for imperfections in data collection.  Most current pulsar
  detection systems use a three-level correlarator with $\beta_0$ of
  1.16 (see LK), and this is the value we shall use.  For $n_{p}$, the
  number of polarizations recorded and summed in the detection
  process, we will use $n_p=2$ because  typically two
  polarizations are summed during pulsar detection scans.
  ${\rm SNR}_{\rm min}$ is the minimum detectable signal-to-noise ratio
  required in a search; we will take this to be 5. 
%
%
%
For the time pointed at the source,
  $t_{obs}$, we will assume a 1 hour observing time.  
  The bandwidth of the recorded data $\Delta f$ depends highly on the
  pulsar detection instruments used at a particular
  telescope. Bandwidths typically range from 100MHz to 800MHz.
  $S_{sys}$ is the system equivalent flux density which depends
  strongly on the collecting area of the telescope and the raw antenna
  sensitivity (see LK).
  The relevant characteristics of 
  current existing radio telescopes and of possible future radio
  telescopes are detailed in Table~\ref{tab:tele}. 
\begin{table}[h!b!p!]
  \begin{center}
  \begin{tabular}{ | c | c | c | c | }
  \hline
  Telescope Name &$S_{\rm sys}$(Jy) &$\Delta f({\rm MHz})$ &$S_{\rm min}$(mJy) \\
  \hline Parkes & 30 & 340 & .022 \\
         GBT& 10 & 800 & .0048 \\
         FAST& 1.5 & 800 & .00072 \\
         SKA& .23 & 800 & .00011 \\
  \hline
  \end{tabular}
  \end{center}
  \caption{Theoretical $S_{\rm min}$  values for
two existing radio telescopes (Parkes, GBT) and two
  planned radio telescopes (FAST, SKA). The $S_{\rm sys}$ values for the
existing radio telescopes are measured system equivalent flux
  densities and the $\Delta f$ values are for current recording
instruments. The $S_{\rm sys}$ values for the two planned telescopes are
  theoretical values and the $\Delta f$ values were chosen to be the
same as the best existing recording instruments.}
  \label{tab:tele}
  \end{table}

  \maketitle
  \section{Probability calculations}\label{sec:prob}

In this section we show how to calculate the probability that
radiation from a single pulsar is detectable by a telescope on Earth,
after having passed through the strong-field region of the black hole.
This calculation naturally breaks down into two parts: determining
what orientations of the pulsar and black hole produce strongly-bent
beams, and determining where the Earth must be positioned relative to
the system in order to detect those beams.

In Paper~II it was shown that for any relative position of pulsar,
black hole, and Earth, there are a set of directions, called
``keyholes,'' in which a photon could be emitted from the pulsar, pass
around the black hole, and arrive at the Earth.  These keyholes are
typically within a few Schwarzschild radii of the black hole, so when
the pulsar is far (many Schwarzschild radii) from the black hole, we
can treat the keyhole as co-located with the black hole: that is, the
pulsar beam must sweep across the black hole.  The first part of the
probability calculation is to determine for what fraction of the
pulsar's orbit it is in a position to illuminate the black hole with
its beam.

\begin{figure}
\begin{center}
\resizebox{6.5in}{!}{\includegraphics{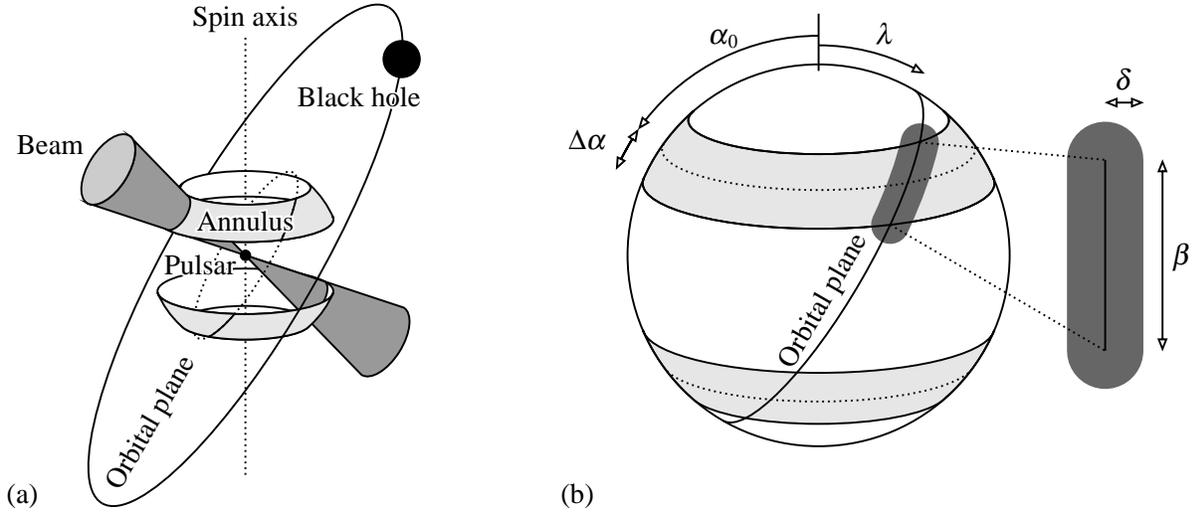}}
\end{center}
\caption{The geometry of a pulsar-black hole
system as seen from the pulsar's reference frame. Panel (a) shows how the
pulsar beam sweeps out an annular region in the pulsar's sky, which
may intersect with the orbital plane of the pulsar-black hole system.
Panel (b) shows the pulsar sky and labels the angles describing the
orientation: $\lambda$ is the angle between the spin axis and orbital
plane, $\alpha_0$ the angle between the spin axis and the center of the
illuminated annulus (or beam), and $\Delta\alpha$ is the half-width of
the annulus (or beam).  The zoom on the right shows the region in the
pulsar's sky in which the Earth can lie if the photons reaching the
Earth from the strong-field regime are to be bent by no more than an
angle $\delta$.}\label{fig:annulus}
\end{figure}

We view the system from the perspective of the pulsar, so that the
black hole traverses the sky of the pulsar along a great circle
corresponding to the orbital plane. (This great circle in the pulsar
sky does {\rm not} imply that the pulsar-SMBH distance is constant.)
Meanwhile, the pulsar spins about its rotation axis, and emits
radiation in a cone offset from that axis: once per pulsar rotation
the cone sweeps out an annulus in the sky of the pulsar.  This is
illustrated in Fig.~\ref{fig:annulus}, where $\alpha_0$,
$\Delta\alpha$ and $\lambda$ have the meaning described in
Sec.~\ref{sec:bg} and pictured in Fig.~\ref{fig:model}.

\begin{figure}
\begin{center}
\resizebox{3.25in}{!}{\includegraphics{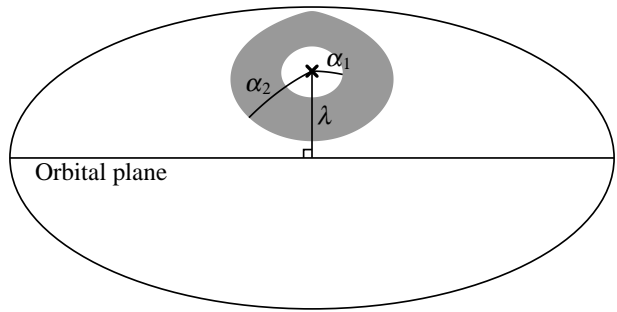}}
\end{center}
\caption{Schematic of the pulsar sky for case~1, in
which the orbital plane does not intersect the annulus.}\label{fig:beta1} 
\end{figure}

\begin{figure}
\begin{center}
\resizebox{2.8in}{!}{\includegraphics{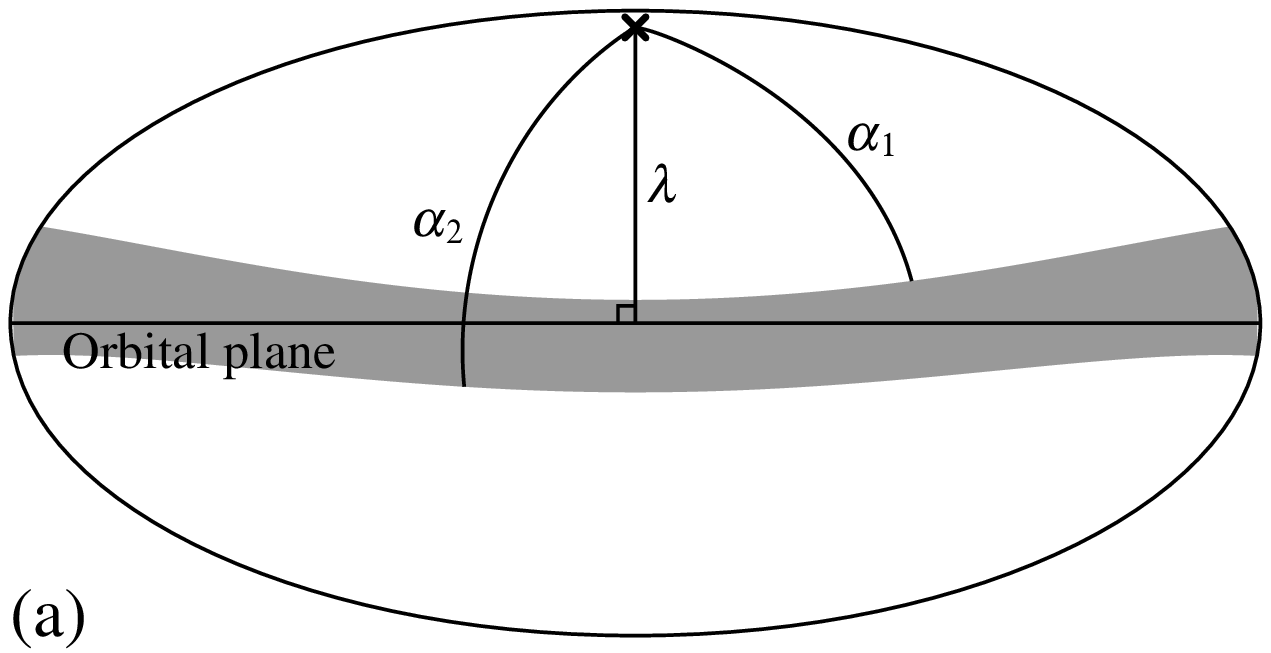}}\hfill
\resizebox{2.8in}{!}{\includegraphics{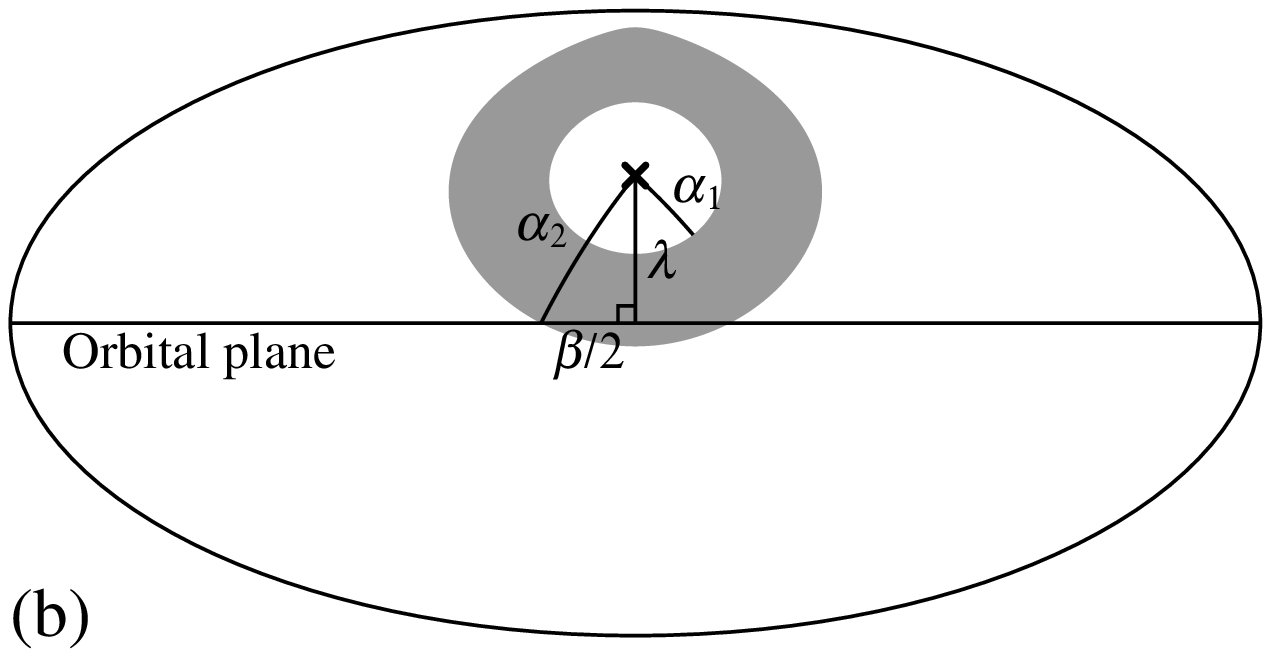}}
\end{center}
\caption{Schematic of pulsar sky for case~2, in
which the orbital plane passes between the outer and inner edges of
the annulus.  In panel~(a), the orbital plane is entirely contained
within the annulus.  In panel~(b), it crosses the outer edge of the
annulus at two points symmetric about the meridian containing the spin
axis.}\label{fig:beta2} 
\end{figure}

\begin{figure}
\begin{center}
\resizebox{3.25in}{!}{\includegraphics{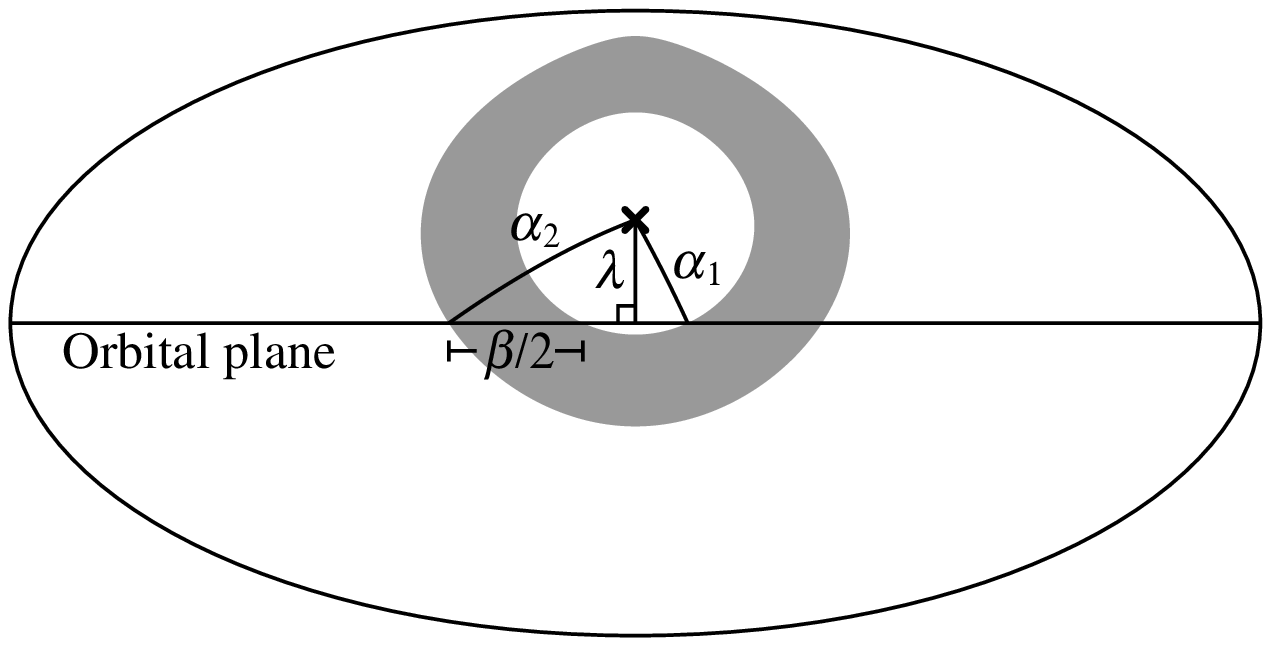}}
\end{center}
\caption{Schematic of pulsar sky for case~3, in
which orbital plane intesects both the inner and outer edges of the
annulus.}\label{fig:beta3} 
\end{figure}

With Fig.~\ref{fig:annulus}(b) we introduce the angle $\beta$,
the total arc length (if any) over which the annulus intersects the
orbital plane.
When calculating $\beta$, it is useful to focus on one hemisphere at a
time and to label the two edges of the pulsar's radiation cone.  We
will define these two edges as $\alpha_1=\alpha_0-\Delta\alpha$ and
$\alpha_2=\alpha_0+\Delta\alpha$.  We break the calculation of $\beta$
into three cases, with one case having two subcases.  The first case
is that in which the orbital plane of the system is
never illuminated by the pulsar's radiation, $\alpha_2<\lambda$, as
shown in Fig.~\ref{fig:beta1}.  In this case, the value of $\beta$
is trivially zero.
The second case is the case in which the orbital plane passes between
the outer and inner edges of the annulus:
$\alpha_1\leq\lambda<\alpha_2$.  This case has two subcases. In the first
subcase, that for $\alpha_2>\pi/2$,
the entire orbital plane is illuminated by the
pulsar, as
shown in Fig.~\ref{fig:beta2}(a),
  The value of $\beta$ in this case is the entire
range $\pi$ that lies in that hemisphere of the pulsar's sky. (Remember
that we are assuming symmetric emission about the pulsar's rotational
plane, so that the two beams together illuminate the full $2\pi$ range
of the orbital plane.)
The second subcase, when $\alpha_2\leq\pi/2$, has the outer
edge of the annulus intersecting the orbital plane twice, as shown in
Figure~\ref{fig:beta2}(b).  In this case, $\beta$ can be found from
the spherical triangle version of Pythagoras's theorem, applied to the
triangle with hypotenuse $\alpha_2$ and sides $\lambda$ and $\beta/2$:
$\cos\alpha_2=\cos\lambda\cos\beta/2$, whence
$\beta=2\arccos(\cos\alpha_2/\cos\lambda)$.
The final case has both the outer and inner edges of the annulus
crossing the orbital plane, $\alpha_1>\lambda$, as illustrated in
Figure~\ref{fig:beta3}.  The calculation proceeds as in the previous
case, but considers only the range of $\beta$ \emph{between} the
triangles with hypotenuses $\alpha_2$ and $\alpha_1$:
$\beta=2\arccos(\cos\alpha_2/\cos\lambda) -
2\arccos(\cos\alpha_1/\cos\lambda)$.
The cases and calculations for $\beta$ are summarized in the following:
$$
\begin{array}{|c|cc|l|}
\hline
\mbox{Case} & \multicolumn{2}{|c|}{\mbox{Condition}} &
~\mbox{$\beta$ Calculation}\\
\hline & & & \\[-2ex]
\begin{array}{l}1~\end{array} & \alpha_2\leq\lambda & &
\begin{array}{l}\beta=0\end{array} \\[1ex]
\begin{array}{l}\mbox{2a}\\[1ex]\mbox{2b}\end{array} &
\alpha_1\leq\lambda<\alpha_2 &
\left\{\begin{array}{c} \alpha_2\geq\pi/2 \\[1ex] \alpha_2<\pi/2
\end{array}\right. &

\begin{array}{l} \beta=\pi \\[1ex]
\beta=2\arccos\left(\frac{\cos\alpha_2}{\cos\lambda}\right) \end{array}\\[3ex]
\begin{array}{l}3~\end{array} & \lambda<\alpha_1 & &
\begin{array}{l}
\beta=2\arccos\left(\frac{\cos\alpha_2}{\cos\lambda}\right) - 
2\arccos\left(\frac{\cos\alpha_1}{\cos\lambda}\right)\end{array} \\[1ex]
\hline
\end{array}
$$

The second part of the probability calculation is to determine how
often the Earth will be in a position to detect strongly-bent beams
from the system. This imposes geometric limitations to ensure that we
are considering beams that are significantly deflected by the black
hole, but that are not deflected so strongly that they are attenuated
to a flux too low to be detected. These constraints turn out to place
limits on the acceptable range of $\phi_{\rm out}$.

\begin{figure}[hb]
\begin{center}
\includegraphics[width=.8\textwidth ]{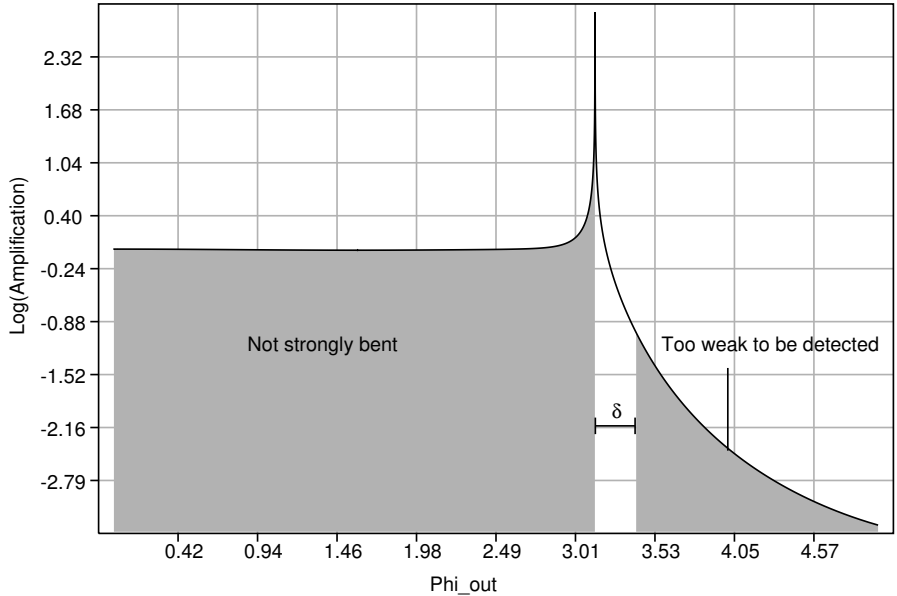}
\caption{The intensity ratio ${I}/{I_0}$ as a function of $\phi_{\rm out}$, 
for $r_0=100M$.
The shaded regions
show the ranges of 
$\phi_{\rm out}$, 
 that are not of interest either because the pulsar beam 
is too weakly deflected or because it is too strongly 
attenuated.}
\label{fig:ampvphiout}
\end{center}
\end{figure}

The determination of that range is shown in Fig.~\ref{fig:ampvphiout}
for the case $r_0=100M$.  That figure shows the dramatic amplification
at $\phi_{\rm out}=\pi$, corresponding to strong lensing.  For
$\phi_{\rm out}$ slightly less than $\pi$, the attenuation factor is
unity, and the bending is not significant. There {\em is} bending for
$\phi_{\rm out}<\pi$, but the range of 
$\phi_{\rm out}$ for which there is significant bending is small. Moreover, 
this range is 
even smaller than in Fig.~\ref{fig:ampvphiout} for the larger, more
relevant values of $r_0/M$. As a convenient approximation, therefore,
we will consider ``strong bending'' only for $\phi_{\rm out}\geq\pi$.
The figure shows that as $\phi_{\rm out}$ incresses beyond $\pi$ the
attenuation becomes greater and greater.

Our approach will be to specify a radius of emission $r_0$ and a
minimum acceptable value of $I/I_0$. From a calculation like that
shown in Fig.~\ref{fig:ampvphiout} we then find the value of the angle
$\delta$, the value of $\phi_{\rm out}-\pi$ at which the attenuation
is that of the minimum acceptable value of $I/I_0$.  This value of
$\delta$ determines the range of directions in which the Earth must
located if an Earth telescope is to detect the beam: the Earth must
lie no more than an angle $\delta$ from the pulsar-black hole axis.

The corresponding region on the
pulsar sky is illustrated in Fig.~\ref{fig:annulus}(b).  Since
$\delta$ is typically very small, we can express this area using a
flat-space approximation: $2\beta\delta+\pi\delta^2$.  The probability
that observers on Earth can detect strongly-bent beams from a given
pulsar is given by the size of this area over the angular area of one
hemisphere (again, we assume symmetry across the rotation plane of the
pulsar):
\begin{equation}
P_1 \;\;=\;\; \frac{2\beta\delta+\pi\delta^2}{2\pi}\,.
\label{eq:prob}
\end{equation}
This of course assumes nonzero $\beta$: if $\beta=0$, the black hole
is not illuminated by the pulsar, and there are no strongly-bent
beams, so $P_1=0$.  Note that $\beta$ is a function of the underlying
parameters $\lambda$, $\alpha_0$, and  $\Delta\alpha$, and note that
$\delta$ is a function of the underlying parameters $r_0/M$ and
$I_\mathrm{min}/I_0$.

  \maketitle
  \section{Results}\label{sec:results}

Now that we have shown how to calculate the probability of Earth
detection for a particular pulsar, we will describe how we estimated
the number of pulsars that would be detected given assumptions about
the distributions of pulsar characteristics.  We ran Monte Carlo
simulations that selected $\lambda$ from a uniform distribution of
$\cos{\lambda}$, so that the direction of the pulsar spin axis was
uniformly distributed over the sky; similary $\alpha_0$ was chosen
from a uniform distribution of $\cos{\alpha_0}$. Then a value for the
pulsar's flux, $S$, was chosen from the distribution shown in
Fig.~\ref{fig:fluxgc}. Lastly, a value of $r_0/M$ was chosen from the
distribution in Eq.~(\ref{nvsr0}); we cut this distribution off at
$r_0/M=400,000$ since pulsars beyond that point contribute little to
the total probability (see below).

The simulations took the pulsar parameters ($r_0$, $S$) chosen by the
Monte Carlo method.  From these a determination was made of the minimum
value of $I/I_0$ (equivalently $S_{\rm min}/S$) that can be detected for 
those pulsar parameters. From $I/I_0$ and $r_0$, the value of $\delta$
was determined. The value of $\beta$ was detemined through the calculations
described in the previous section applied to the
values of $\lambda$ and $\alpha_0$ chosen by the Monte Carlo method.  
The justification for using
$\Delta\alpha=9^\circ$ has been explained in Sec.~\ref{sec:bg}. $P_1$ was
then calculated with Eq.~\ref{eq:prob}. This gives us the probability
that a single pulsar's beam will explore the black hole's strong
gravitational field and still be detectable once it reaches the
Earth. Monte Carlo simulations were repeated to insure that results
were consistent to better than 1\%. 
We then multiplied
 our result by the total number of
pulsars out to $r_0=400,000$, according to the Pfahl and Loeb distribution
of Eq.~(\ref{nvsr0}).
The result is the total number of pulsars ($P_{\rm tot}$) that will, at
some point in their orbit about the central SMBH, both illuminate the strong
gravitational field and be detectable at Earth. Table~\ref{tab:ptot}
gives this number for the four telescopes considered in
Sec.~\ref{sec:bg}.  

\begin{table}[h!b!p!]
  \begin{center}
  \begin{tabular}{ | c | c | c |}
  \hline
  Telescope Name &$S_{\rm sys}$(Jy) & $P_{\rm tot}$\\
  \hline Parkes & .022 & 8.75776\\
         GBT& .0048 & 15.8075\\
         FAST& .00072 & 111.677\\
         SKA& .00011 & 207.021\\
  \hline
  \end{tabular}
  \end{center}
  \caption{$S_{\rm min}$ and $P_{\rm tot}$ values for
  two existing radio telescopes (Parkes, GBT) and two
  planned radio telescopes (FAST, SKA).}
  \label{tab:ptot}
  \end{table}

\begin{figure}[hb]
\begin{center}
\includegraphics[width=.4\textwidth ]{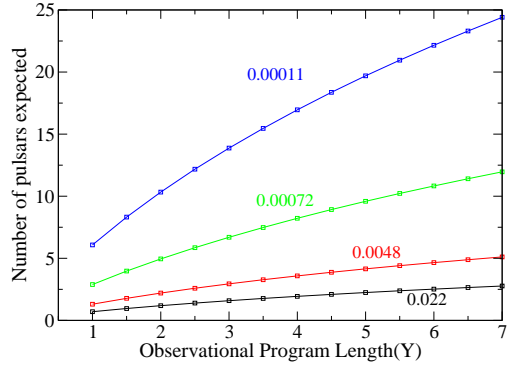}
\caption{Expected number of pulsars observed 
as a function of observational program duration.
Curves are labeled 
with the values of $S_{\rm min}$, in mJy, assumed in the computations.}
\label{fig:obsvsT}
\end{center}
\end{figure}

\begin{figure}[hb]
\begin{center}
\includegraphics[width=.4\textwidth ]{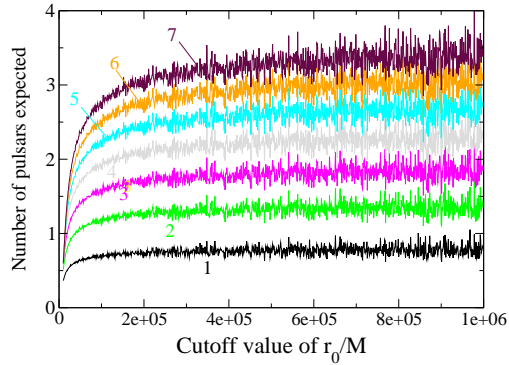}
\caption{The number of pulsars expected, for $S_{\rm min}=0.022$\,mJy,
  as a function of the cutoff $r_0/M$ for the distribution in
  Eq.~(\ref{nvsr0}). Each curve is labeled with the number of years
  assumed for the search.}
\label{fig:numpulvsmaxr0}
\end{center}
\end{figure}

The number of pulsars that are observable at some point in their orbit
is not directly relevant if the orbital time is much larger than the
duration of an observing program. For that reason we introduce a more
useful number, the ``observability,'' $P_{\rm obs}$, to describe the
probability of observing a given pulsar in a limited-time observing
program.  In order to calculate this number, we ran the Monte Carlo
simulations with a specified observational program duration
($L$). Once $P_1$ was calculated, we then compared the pulsar's orbital
period ($T$) to $L$.  If  $T$ was less than or equal to $L$, then
$P_{\rm obs}$ was taken to be $P_1$. If $T$ was greater than $L$, then
$P_1$ was replaced by 
\begin{equation}
P_{\rm obs}=\frac{P_1\times{L}}{T}\,,
\label{eq:observability}
\end{equation}
and the result was multiplied by the total number of
pulsars. Figure~\ref{fig:obsvsT} shows the results of the Monte Carlo
simulations for observational program durations ranging from one year to
seven years.
Figure~\ref{fig:numpulvsmaxr0} shows the variation in the number of
pulsars detected, for $S_{\rm min}=0.022$\,mJY, as the cutoff radius
is changed, and justifies our use of the cutoff at $r_0/M=4\times10^5$.

  \maketitle
  \section{Observing Programs and Strategies}\label{sec:obsstrat}

A natural first question about observing strongly deflected beams is
``how will we know that they are strongly deflected?''  The answer
starts with the fact that the angle through which the beam is
``strongly'' deflected is not large. For our paradigmatic case,
$r_0=10^4M$, the bending is approximately 0.036 rad. For larger $r_0$
the deflection, for a given $S_{\rm min}$, is even smaller.

\begin{figure}[hf]
  \begin{center}
  \includegraphics[width=.4\textwidth ]{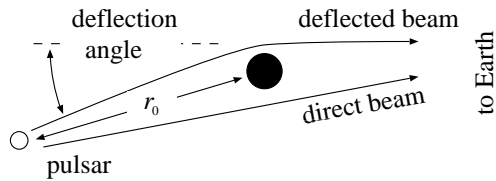}
  \caption{The geometry of the observation of direct and deflected 
beams.}
  \label{fig:directdeflected}
  \end{center}
  \end{figure}

Since the ``strong'' deflection is small, we will receive a deflected
beam only when the emitting pulsar, the SMBH and the Earth are almost
on a straight line. Since  pulsar beam widths are large
compared to the deflection, this means that if the Earth receives 
the deflected beam, it will also receive the direct beam. The geometry of 
the direct and deflected beams is shown in Fig.~\ref{fig:directdeflected}, where
we see that the angle, at reception, of the direct and deflected beams is
not of order $r_0$ divided by the Earth-SMBH distance (8\,kpc), but rather 
is of order of that number multiplied by the deflection angle. The result,
an angle of order $10^{-8}$ radians, is less than the resolution of radio
telescopes. We conclude that any monitoring of the innermost region of Sgr~A$^*$
for a deflected beam, will also monitor for a direct beam. 

The criterion for the detection of a strongly deflected beam will be
that it is one of a pair of pulses detected with very similar pulse
periods, differing only due to a phase modulation caused by variations
in the propagation times along the two paths: the two sets of pulses
will differ in pulse period by a fractional amount of order the pulsar
velocity divided by $c$. (See Papers~I and~II for more detail on phase effects 
and intensity effects of deflection.)
If a deflected beam is detected we therefore assume that it will be relatively
simple to identify it as deflected.

If the Pfahl and Loeb distribution of Eq.~(\ref{nvsr0}) is
approximately valid, the results of the previous section, especially
Fig.~\ref{fig:obsvsT}, suggest that there is a very good
chance of observing a strongly deflected pulsar beam with an observing
program of 3-5 years, using existing telescopes.  Monitoring of
Sgr~A$^*$, of course, will face the problem of dispersion by the
plasma density in the Galactic center.  Assuming dispersion measures
in excess of $1000\,\mathrm{pc}/\mathrm{cm}^3$, searches would have to
be conducted in S-band (2--4\,GHz) or higher frequencies in order to
resolve pulses, rather than the more conventional L-band.

We note that the pulsar fluxes and current telescope sensitivities
were computed for the L band, since that is where most pulsar
luminosities have been measured.  Relative sensitivities in the S band
may be somewhat lower.  However, we note that the baseline value of
$S_{\rm min}=0.022$\,mJy for the Parkes telescope is very conservative
in view of the sensitivities that will soon be available. See
Table~\ref{tab:tele}, and the corresponding estimates of detectable
pulsars in Fig.~\ref{fig:obsvsT}.  Even with the current estimated
senstitivity of Parkes there is a good chance of detecting
strongly-bent pulsar beams in a 2--3~year observing program; with more
sensitive telescopes such as GBT or FAST (under development), the
number could rise to several pulsars.  Longer observing times increase
the number of detectable pulsars by allowing long-period pulsars more
time to come into alignment.

An appropriate multiyear observing program can be carried out ``in
background'' at a telescope. Observations must be made sufficiently
frequently not to miss the relatively short epoch during which the
pulsar/SMBH/Earth alignment leads to a strongly deflected beam meeting
the $I/I_0$ criterion. That epoch is of order of the orbital time
multiplied by the ratio $\delta/(2\pi)$. For our
prototypical choice $r_0=10^4\,M$, this epoch is on the order of a
week. Searches for deflected beams in Sgr~A$^*$, therefore, would have
to be made every other day. The observing session would be of a
duration of that used for any other pulsar search, on the order of 1
hour. (Longer observations might be considered, in view of the effect
shown in Eq.~(\ref{eq:minflux}) of $t_{obs}$ on $S_{\rm min}$, and hence
on the $I/I_0$ cutoff.

An important question to ask is why we do not yet have evidence of the
assumed large population of pulsars in Sgr~A$^*$. This can be explained
by the lack of any concerted effort to search the Galactic center for
pulsars, and the need for S band searches with large dispersion corrections.  In addition the Galactic
Center is out of view for the Arecibo telescope, and in a portion of the
sky with reduced sensitivity for GBT.
Currently, Parkes is the largest telescope with a clear view of the
Galactic Center; the FAST telescope will substantially increase our
sensitivity to the Galactic Center.

It should be noted that, at least to some extent, monitoring the Galactic
center for strongly deflected beam would constitute a more general search 
for pulsars in that region.  The possibility of detecting deflected beams
and making measurements of the parameters of the SMBH provides added
scientific motivation for such a survey.

  \section{Conclusions}\label{sec:conc}
 
  Our estimates suggest that a multi-year program that monitors
  Sgr~A$^*$ with radio observations for one hour every other day has a
  reasonable probability of detecting pulsar beams that have been
  strongly deflected by our Galaxy's SMBH. With instruments coming in
  the near future, in particular, FAST and SKA, the probability should
  become high enough so that a three year observational program either
  detects a strongly deflected beam, or that the failure to make such
  a detection puts useful limits on the density of pulsars in the
  Galactic Center.

  Our estimates in this paper constitute a first step in the study of
  probabilities of detection of a strongly deflected beam.  The
  intention was to establish whether the probabilities are so small
  that observations are out of the question, or so large that current
  observations rule out models, like that of
  \cite{2004ApJ...615..253P}, with a significant density of pulsars
  in Sgr~A$^*$. The estimates in this paper establish neither extreme:
  a concerted observing program with the best current telescopes would
  optimistically detect only a few such pulsars.  This provides motivation
  for such a program, and also for further study of the problem of
  pulsar beam deflection by SMBHs.

Such an improved study would have to include effects of spin of the
SMBH, and of eccentricity of orbits. While our approach of using
averages and simple assumptions was appropriate to the purpose of this
paper, effects due to SMBH spin, and high eccentricity, could increase
the parameter space of pulsar configurations whose beams can reach the
Earth.  Such work is now underway.

The most exciting result of these preliminary estimates is their
indication that we are potentially on the verge of detecting a new
phenomenon: pulsar beams that have passed through the strong field
region of the SMBH at the center of our Galaxy, beams that can bring
us information about the properties of the SMBH and its surrounding
spacetime might be inaccessible in any other way.

We gratefully acknowledge support by the National Science Foundation under grants
AST0545837, PHY0554367 and HRD0734800. We also thank the Center for Gravitational
Wave Astronomy at the University of Texas at Brownsville.

\section{Appendix}

We are primarily interested in values of $\phi_{\rm in}$  that are only slightly
smaller than $\pi$.  In this case the photon path starting at some very large 
radius will penetrate to small radii, and almost all the bending 
will take place at small radii. We can then find the bending for emission from 
the very large radius, say $R_0=10,000\,M$,  by considering  the bending only 
interior to a smaller large radius, say  $r_0=100\,M$. In effect, we are considering
the large radius region from $R_{\rm o}$ to $r_{\rm o}$ to be flat space.
 From the 
the curve $F(\phi_{\rm in};r)$ for $r_0=100\,M$, therefore, we can infer the 
curve for  $R_0=10,000\,M$, and for all larger radii (provided, of course,
that $\phi_{\rm in}$ is near $\pi$ so that both $R_{\rm o}$ and $r_{\rm o}$ 
are much larger than the radii at which the bending occurs).
\begin{figure}[hf]
  \begin{center}
  \includegraphics[width=.4\textwidth ]{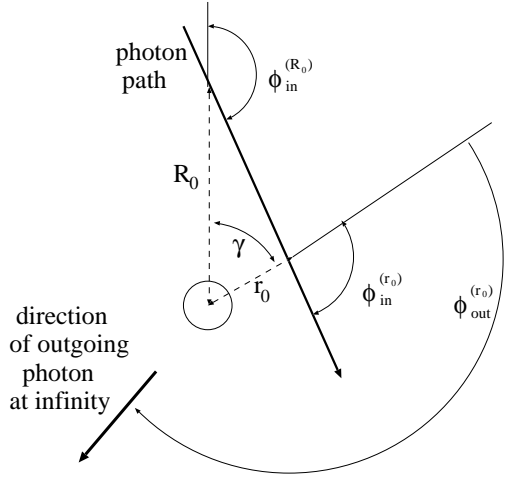}
  \caption{The flat spacetime propagation from large radius to very
    large radius.  Angles $\phi_{\rm in}$ for the large, and the very
    large emission radius are distinguished with the superscripts
    $(r_0)$ and $(R_0)$, respectively.  }
  \label{fig:photonpath}
  \end{center}
  \end{figure}

  To find the $\phi_{ \rm out}^{\rm( R_0)}$ for a photon emitted at a
  very large $R_0$, we choose the  smaller $r_0$ along the  future path of
the photon to be a radius for which we know the
curve $\phi_{\rm out}^{\rm(r_0)}=  F(\phi_{\rm in}^{\rm(r_0)};r_0)$. 
Here superscripts $(r_0)$ and $(R_0)$  distinguish the angles associated
with the two radii.
 The Euclidean
geometry  relating 
$\phi_{\rm in}^{\rm(r_0)}$ and $\phi_{\rm in}^{\rm(R_0)}$
is given by the law of sines to be
\begin{equation}\label{phinRin}
  \phi_{\rm in}^{\rm({r_{\rm 0}})}=\sin^{-1}\!\left([R_0/r_0]\sin{\phi_{\rm in}^{\rm({R}_{\rm 0}) }}\right)\,,
\end{equation}
and we choose the branch of $\sin^{-1}$ so that
$\phi_{\rm in}^{\rm({r_{\rm 0}})}>\pi/2$.

\begin{figure}[hf]
  \begin{center}
  \includegraphics[width=.55\textwidth ]{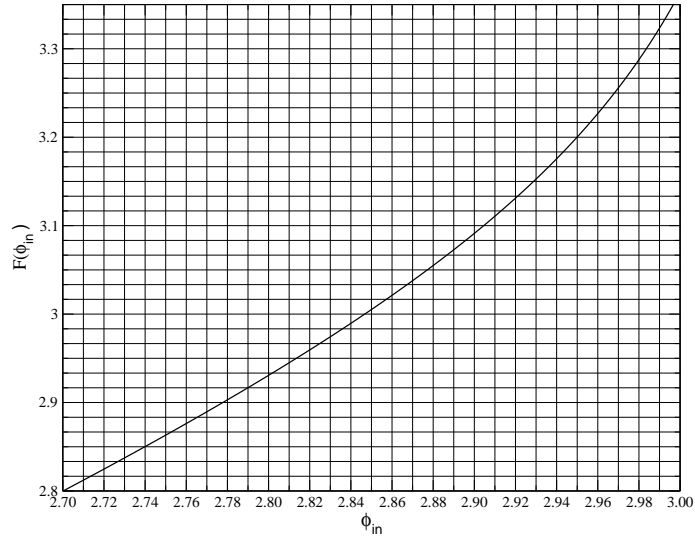}
  \caption{The bending function $\phi_{\rm out}=F(\phi_{\rm in})$ for
  bending angles near $\pi$, for an initial radius $r_0=100M$.}
  \label{fig:r100}
  \end{center}
  \end{figure}

We next notice that $\phi_{\rm out}$ for the photon starting at $r_0$ is less than 
$\phi_{\rm out}$ for that same photon world line considered to start at 
$R_0$, in flat space, according to 
\begin{equation}
  \phi_{\rm out}^{(\rm{R}_{0})}=\phi_{\rm out}^{(\rm{r}_{0})}+\gamma\,.
\end{equation}
With $\gamma$ evaluated in terms of the ingoing angles, this becomes
\begin{equation}\label{phinRout}
  \phi_{\rm out}^{\rm({R}_{0})}=   \phi_{\rm out}^{\rm({r}_{0})}+\phi_{\rm in}^{\rm({R}_{0})}-\phi_{\rm in}^{\rm({r}_{0})}\,.
\end{equation}
From a combination of Eqs.~(\ref{phinRin}) and~(\ref{phinRout}),
written in terms of the function $F(\phi;r)$, the full expression can be given
as:
\begin{equation}
F(\phi;R_0) \;\;=\;\; \phi + F\left[\sin^{-1}\!\left(\mbox{$\frac{R_0}{r_0}$}
\sin\phi\right);r_0\right] - \sin^{-1}\!\left(\mbox{$\frac{R_0}{r_0}$}
\sin\phi\right)
\end{equation}
Thus, knowing the $F$ function for any (sufficiently large) $r_0$, we
can evaluate it for any larger $R_0$, and thus determine the maximum
deflection angle $\delta$ via Eq.~(\ref{eq:ampfactor}).
Figure~\ref{fig:r100} shows our reference bending function for a
radius $r_0=100M$.


\begin{thebibliography}{15}
\expandafter\ifx\csname natexlab\endcsname\relax\def\natexlab#1{#1}\fi

\bibitem[{{ATNF Pulsar Catalog}(2010)}]{ATNF}
{ATNF Pulsar Catalog}. 2010, http://www.atnf.csiro.au/research/pulsar/psrcat

\bibitem[{{Deegan} \& {Nayakshin}(2007)}]{2007MNRAS.377..897D}
{Deegan}, P., \& {Nayakshin}, S. 2007, \mnras, 377, 897

\bibitem[{{Eisenhauer} {et~al.}(2005){Eisenhauer}, {Genzel}, {Alexander},
  {Abuter}, {Paumard}, {Ott}, {Gilbert}, {Gillessen}, {Horrobin}, {Trippe},
  {Bonnet}, {Dumas}, {Hubin}, {Kaufer}, {Kissler-Patig}, {Monnet},
  {Str{\"o}bele}, {Szeifert}, {Eckart}, {Sch{\"o}del}, \&
  {Zucker}}]{2005ApJ...628..246E}
{Eisenhauer}, F., {Genzel}, R., {Alexander}, T., {Abuter}, R., {Paumard}, T.,
  {Ott}, T., {Gilbert}, A., {Gillessen}, S., {Horrobin}, M., {Trippe}, S.,
  {Bonnet}, H., {Dumas}, C., {Hubin}, N., {Kaufer}, A., {Kissler-Patig}, M.,
  {Monnet}, G., {Str{\"o}bele}, S., {Szeifert}, T., {Eckart}, A.,
  {Sch{\"o}del}, R., \& {Zucker}, S. 2005, \apj, 628, 246

\bibitem[{{Freitag} {et~al.}(2006){Freitag}, {Amaro-Seoane}, \&
  {Kalogera}}]{2006JPhCS..54..252F}
{Freitag}, M., {Amaro-Seoane}, P., \& {Kalogera}, V. 2006, Journal of Physics
  Conference Series, 54, 252

\bibitem[{{Genzel} {et~al.}(2003){Genzel}, {Sch{\"o}del}, {Ott}, {Eckart},
  {Alexander}, {Lacombe}, {Rouan}, \& {Aschenbach}}]{2003Natur.425..934G}
{Genzel}, R., {Sch{\"o}del}, R., {Ott}, T., {Eckart}, A., {Alexander}, T.,
  {Lacombe}, F., {Rouan}, D., \& {Aschenbach}, B. 2003, \nat, 425, 934

\bibitem[{{Ghez} {et~al.}(2005){Ghez}, {Salim}, {Hornstein}, {Tanner}, {Lu},
  {Morris}, {Becklin}, \& {Duch{\^e}ne}}]{2005ApJ...620..744G}
{Ghez}, A.~M., {Salim}, S., {Hornstein}, S.~D., {Tanner}, A., {Lu}, J.~R.,
  {Morris}, M., {Becklin}, E.~E., \& {Duch{\^e}ne}, G. 2005, \apj, 620, 744

\bibitem[{{Hopman} \& {Alexander}(2006)}]{2006JPhCS..54..321H}
{Hopman}, C., \& {Alexander}, T. 2006, Journal of Physics Conference Series,
  54, 321

\bibitem[{{Lorimer} \& {Kramer}(2004)}]{2004hpa..book.....L}
{Lorimer}, D.~R., \& {Kramer}, M. 2004, {Handbook of Pulsar Astronomy}, ed.
  {Lorimer, D.~R.~\& Kramer, M.}, (LK)

\bibitem[{{Maness} {et~al.}(2007){Maness}, {Martins}, {Trippe}, {Genzel},
  {Graham}, {Sheehy}, {Salaris}, {Gillessen}, {Alexander}, {Paumard}, {Ott},
  {Abuter}, \& {Eisenhauer}}]{2007ApJ...669.1024M}
{Maness}, H., {Martins}, F., {Trippe}, S., {Genzel}, R., {Graham}, J.~R.,
  {Sheehy}, C., {Salaris}, M., {Gillessen}, S., {Alexander}, T., {Paumard}, T.,
  {Ott}, T., {Abuter}, R., \& {Eisenhauer}, F. 2007, \apj, 669, 1024

\bibitem[{{Melia} {et~al.}(2001){Melia}, {Bromley}, {Liu}, \&
  {Walker}}]{MeliaBromleyEtal}
{Melia}, F., {Bromley}, B.~C., {Liu}, S., \& {Walker}, C.~K. 2001, \apjl, 554,
  L37

\bibitem[{{Muno} {et~al.}(2005){Muno}, {Pfahl}, {Baganoff}, {Brandt}, {Ghez},
  {Lu}, \& {Morris}}]{2005ApJ...622L.113M}
{Muno}, M.~P., {Pfahl}, E., {Baganoff}, F.~K., {Brandt}, W.~N., {Ghez}, A.,
  {Lu}, J., \& {Morris}, M.~R. 2005, \apjl, 622, L113

\bibitem[{{Nayakshin} \& {Sunyaev}(2005)}]{2005MNRAS.364L..23N}
{Nayakshin}, S., \& {Sunyaev}, R. 2005, \mnras, 364, L23

\bibitem[{{Pfahl} \& {Loeb}(2004)}]{2004ApJ...615..253P}
{Pfahl}, E., \& {Loeb}, A. 2004, \apj, 615, 253

\bibitem[{{Wang} {et~al.}(2009{\natexlab{a}}){Wang}, {Creighton}, {Price}, \&
  {Jenet}}]{2009ApJ...705.1252W}
{Wang}, Y., {Creighton}, T., {Price}, R.~H., \& {Jenet}, F.~A.
  2009{\natexlab{a}}, \apj, 705, 1252, (Paper II)

\bibitem[{{Wang} {et~al.}(2009{\natexlab{b}}){Wang}, {Jenet}, {Creighton}, \&
  {Price}}]{2009ApJ...697..237W}
{Wang}, Y., {Jenet}, F.~A., {Creighton}, T., \& {Price}, R.~H.
  2009{\natexlab{b}}, \apj, 697, 237, (Paper I)

\end{thebibliography}

\end{document}